\documentclass{jpsj3}

\usepackage{bm}

\newcommand{\Tc}{T_{\mbox{\scriptsize C}}}
\newcommand{\TN}{T_{\mbox{\scriptsize N}}}
\newcommand{\mb}{\mu_{\mbox{\scriptsize B}}}

\title{
Novel Magnetic Chiral Structures and Unusual Temperature Hysteresis \\ in the Metallic Helimagnet MnP
}

\author{Teruo Yamazaki$^{1,2}$\thanks{E-mail: t.yamazaki@rs.tus.ac.jp}, Yoshikazu Tabata$^{2}$, Takeshi Waki$^{2}$, Taku J. Sato$^{1 \dag}$, 
Masato Matsuura$^{3 \ddag}$, Kenji Ohoyama$^{3}$, Makoto Yokoyama$^{4}$, and  Hiroyuki Nakamura$^{2}$
}
\inst{
$^{1}$Neutron Science Laboratory, Institute for Solid State Physics, University of Tokyo, Tokai, Ibaraki 319-1106, Japan\\
$^{2}$Department of Materials Science and Engineering, Kyoto University, Kyoto 606-8501, Japan\\
$^{3}$Institute for Materials Research, Tohoku University, Sendai 980-8577, Japan\\
$^{4}$Faculty of Science, Ibaraki University, Ibaraki 310-8512, Japan\\
$*$ Present address: Department of Physics, Faculty of Science and Technology, Tokyo University of Science, Chiba 278-8510, Japan\\ 
$\dag$ Present address: Institute of Multidisciplinary Research for Advanced Materials, Tohoku University, Sendai 980-8577, Japan\\
$\ddag$ Present address: Comprehensive Research Organization for Science and Society, Research Center for Neutron Science and Technology (CROSS), Tokai, Ibaraki 319-1106, Japan \\
}
\abst{
We have reinvestigated the magnetic properties of the classical metallic helimagnet MnP by magnetization and neutron scattering experiments.
Our neutron scattering results indicate that the previously reported magnetic structure in the low-temperature (LT) helimagnetic phase ($T<47$\,K) should be modified
 to an alternately tilted helimagnetic structure produced by the Dzyaloshinsky-Moriya interaction.
In the intermediate temperature (IT) range between the LT helimagnetic phase and the high-temperature (HT) ferromagnetic phase along the c-axis, 47\,K $< T <$ 282\,K, we have found a weak ferromagnetic behavior along the $b$-axis.
Surprisingly, the IT weak ferromagnetic phase has two different states, namely, the large magnetization (LM) and small magnetization (SM) states.
The SM state emerges with cooling from the paramagnetic phase above 292\,K via the HT ferromagnetic phase and LM state emerges with warming from the LT helimagnetic phase.
The weak ferromagnetism along the b-axis and the unusual temperature hysteresis in the IT phase can be understood by assuming a spontaneous formation of the stripe structure consisting of alternately arranged HT ferromagnetic and LT helimagnetic domains.
}

\begin{document}
\maketitle

\section{Introduction}

\begin{figure}[t]
\hspace{-5mm}
\includegraphics[width=150mm]{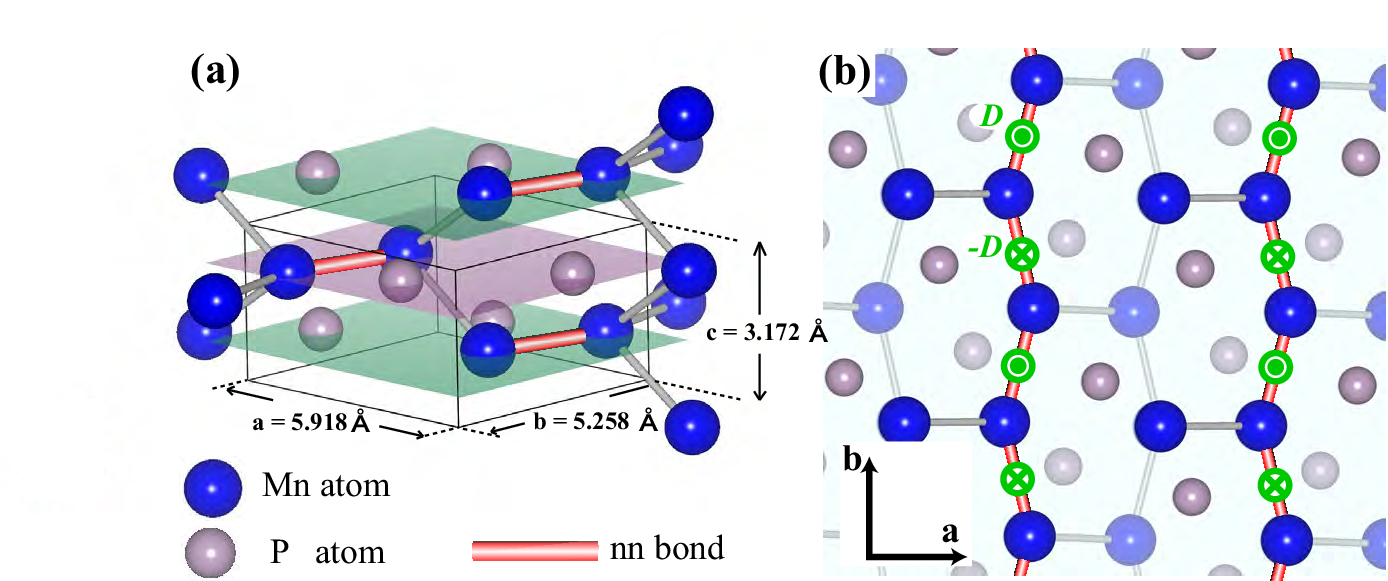}
\caption{(Color online) (a) Schematically illustrated crystal structure of MnP.
The space group is Pbnm.
The thick  lines represent the bonds of nn Mn sites.
(b) Projective figure of the crystal structure of MnP from the c-direction. Possible DM vector at the center of nn Mn sites, directed alternately, are represented by cross and dot marks. }
\label{fig:crystal-1}
\end{figure}

Manganese phosphide (MnP) has been investigated by numerous researchers since the 1960s because it exhibits interesting magnetic properties such as an in-field multistep phase transition~\cite{Huber1964, Forsyth1966, Felcher1966, komatsubara1974, Obara1980, Moon1982}, Lifshits critical behavior \cite{becerra1980, shapira1981, Yoshizawa1985, Bindilatti1989, becerra2000, Zieba2000}, and the magnetocaloric effect \cite{Reis2008}.
The crystal structure is orthorhombic with the space group $Pbnm$ [see Fig.\,\ref{fig:crystal-1}(a)].
The easy axis of the magnetization is the $c$-axis, whereas the $b$- and $a$-axes are the intermediate- and hard-magnetization directions, respectively.
At zero external field, three phase transitions have been reported to date.
The reported phase transitions and corresponding magnetic structures may be summarized as follows.
First, the para-ferromagnetic phase transition takes place at $\Tc= 292$\,K, below which Mn spins are aligned parallel to the $c$-axis~\cite{Huber1964,Felcher1966}.
Becerra has recently reported an additional transition at $T^{\ast} = 282$\,K, immediately below $T_{\rm C}$, suggesting that spins are slightly reoriented toward the $b$-axis below $T^{\ast}$~\cite{becerra2000a}.
At a lower temperature, i.e., $\TN = 47$~K, further transition into a double spiral-type helical structure was reported~\cite{Felcher1966,Forsyth1966,Obara1980}.
In this helimagnetic state, spins rotate in the $bc$-plane (helical plane) with the propagation vector ${\bm Q} = (\delta,0,0)$ normal to the helical plane.
The ordered moment is about 1.3\,$\mb$/Mn atom~\cite{Huber1964,Obara1980}.
In summary, MnP has four phases at zero field: the low-temperature (LT) helimagnetic phase for $T < \TN$, the intermediate-temperature (IT) phase for $\TN  < T < T^{\ast}$, the high-temperature (HT) ferromagnetic phase for $T^{\ast} < T < \Tc$, and the paramagnetic phase for $T > \Tc$.
In our recent study, a divergent behavior of the AC susceptibility along the b-axis has been observed at $T^{\ast}$.
This indicates that $T^{\ast}$ is a second-order magnetic phase transition\cite{Yamazaki2010}. 
Strangely, the anomaly of the AC susceptibility at $T^{\ast}$ strongly depends on the measurement process: 
 the anomaly is strikingly enhanced when the sample is cooled down to the LT phase once. 
This indicates that the LT phase involves a phase transition at $T^{\ast}$, whereas
 the LT helimagnetic state seems to have no connection with the reorient transition at $T^{\ast}$ suggested by Beccera\cite{becerra2000a}.

Thus, in this work, we reviewed magnetic structures of MnP 
 by performing detailed magnetization and neutron scattering measurements, especially aiming to find a relation ship between the LT and IT phases.
As a result, we have observed the following; 
(i) a weak ferromagnetism along the $b$-axis in the IT phase, which is strongly enhanced when the sample is cooled down to the LT phase once, as well as the anomaly of the AC susceptibility at $T^{\ast}$, 
(ii) unusual temperature hysteresis for the weak ferromagnetic moment, 
(iii) a new magnetic reflection at ($\delta$,1,0) in the LT phase ($T < \TN$), which indicates that the helical plane is tilted to the a-direction from the $bc$-plane alternately.
As MnP lacks the inversion symmetry at the center of the nearest-neighbor (nn) Mn bonds, the DM interaction may remain finite between them [see Fig.\,\ref{fig:crystal-1}(b)].
We argued that the tilted helical structure is a chiral structure stabilized by the finite DM interaction in the LT phase.
The weak ferromagnetism and unusual hysteresis are interpreted as a long-periodic stripe structure, 
 consisting of an alternate arrangement of the HT ferromagnetic and LT helimagnetic domains, in the IT phase.
The second-order phase transition at $T^{\ast}$ is a spontaneous formation of a stripe structure.

\section{\label{sec:level2}Experimental Procedure}
The single-crystalline samples of MnP used in this study were grown by the temperature gradient furnace technique\cite{Huber1964}. 
Their magnetization was measured using a superconducting quantum interference device (SQUID) magnetometer (MPMS, Quantum Design) at the Research Center for Low Temperature and Materials Sciences, Kyoto University.
The sizes of the crystals used for the magnetization measurements were $2.5 \times 0.9 \times 0.7$, $0.7 \times 2.5 \times 1.0$, and $0.7 \times 0.6 \times 2.5 \mbox{ mm}^3$ along the $a$-, $b$-, and $c$-axes, respectively.
Single crystal neutron scattering experiments in the $hk0$ scattering plane were performed using the triple-axis spectrometers ISSP-PONTA and ISSP-HER installed at JRR-3 of Japan Atomic Energy Agency (JAEA).
Neutron powder diffraction experiments were performed using the multidetector diffractometer IMR-HERMES  installed at JRR-3~\cite{Ohoyama1998}.
We used a sample with dimensions of $6.1 \times 3.6 \times 9.6 \mbox { mm}^3$ for the single-crystal experiments.
A powder sample with a mass of 3.6 g, obtained by grinding a single-crystalline sample, was used for the powder diffraction experiments.

\section{\label{sec:level3}Experimental Results}
\subsection{\label{sec:level4}Magnetization measurements}

\begin{figure}[h]
\begin{center}
\hspace{-1.5 mm}
\vspace{-8mm}
\includegraphics[width= 96mm]{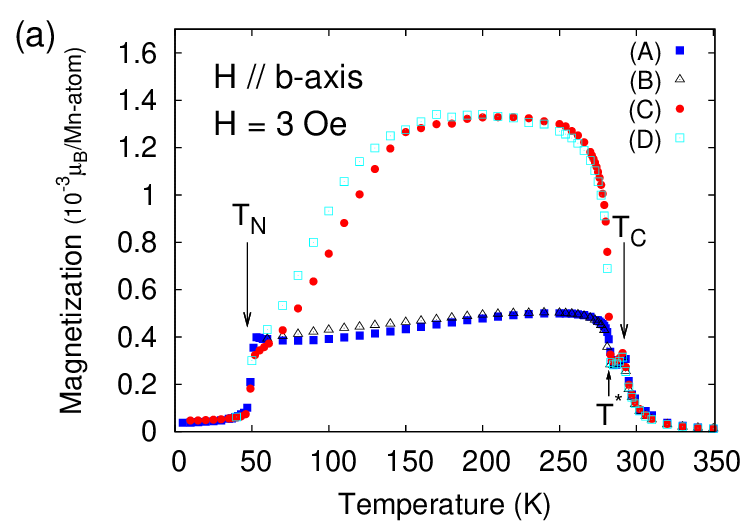}
\includegraphics[width= 89.5mm]{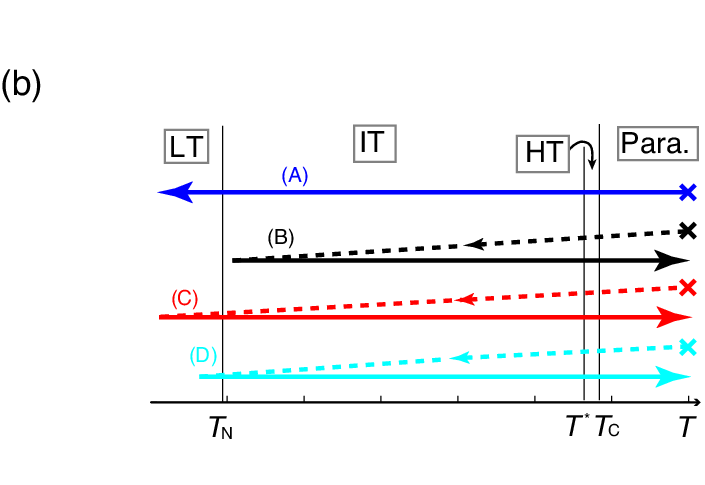}
\caption{(Color online) 
(a) Temperature ($T$) dependences of the magnetization along the $b$-axis of MnP in the various $T$ processes (A)-(D).
(b) Schematic diagram of the measurement processes (A)-(D) and the magnetic phases in MnP, namely, the low-temperature (LT) helimagnetic phase, intermediate-temperature (IT) phase, high-temperature (HT) phase, and paramagnetic phase.
The crosses represent the initial temperature at which a magnetic field of 3\,Oe was applied.
The solid and dotted arrows represent the measurement and approaching procedures for the initial measurement temperatures, respectively.
 }
\label{fig:b-axis} 
\end{center}
\end{figure}

The temperature ($T$) dependences of the magnetization along the $b$-axis in various processes are shown in Fig.~\ref{fig:b-axis}(a).
The measurements were performed using four different $T$ processes
after applying a field of 3\,Oe at 350\,K well above $\Tc$:
 (A) a cool-down process from 350\,K, 
 (B) a warm-up process after cooling the sample down to 55\,K ($> \TN$), 
 and (C) and (D) warm-up processes after cooling the sample down to 5 and 35\,K ($< \TN$) [see Fig.~\ref{fig:b-axis}(b)].
A sharp increase in the magnetization was observed at $T^*$ in all four processes. 
Moreover, a marked dependence of the size of the magnetization in the intermediate phase ($\TN < T < T^{\ast}$) on the $T$ process was observed. 
The magnetizations in the processes (A) and (B) are almost the same.
Similarly, those in the processes (C) and (D) are almost the same.
Surprisingly, the magnitudes of the magnetization in the latter processes, in which the sample has been cooled below $\TN$ once, are about three times larger than those in the former processes, in which the sample has not been cooled below $\TN$.
The unusual temperature hysteresis observed indicates that
  the magnetization along the $b$-axis in the IT phase is strongly enhanced after the sample has undergone a low-temperature helimagnetic phase once.
In contrast, the magnetizations in all the processes collapse in the LT phase ($T < \TN$), HT phase ($T^{\ast} < T < \Tc$), and paramagnetic phase ($T > \Tc$). 

\begin{figure}[t]
\begin{center}
\includegraphics[width= 96mm]{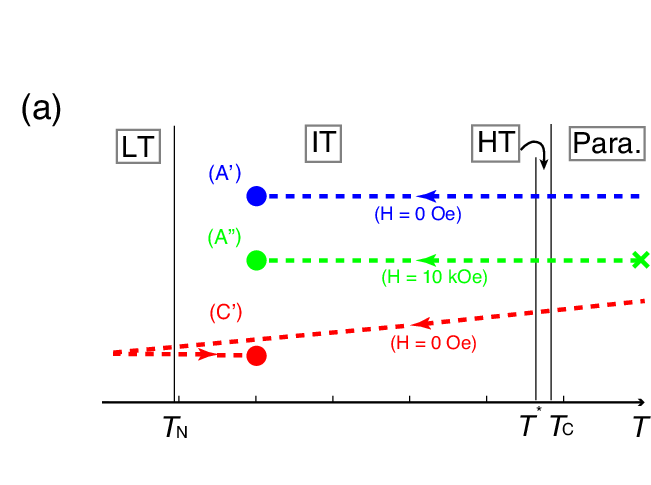}\\
\includegraphics[width=100mm]{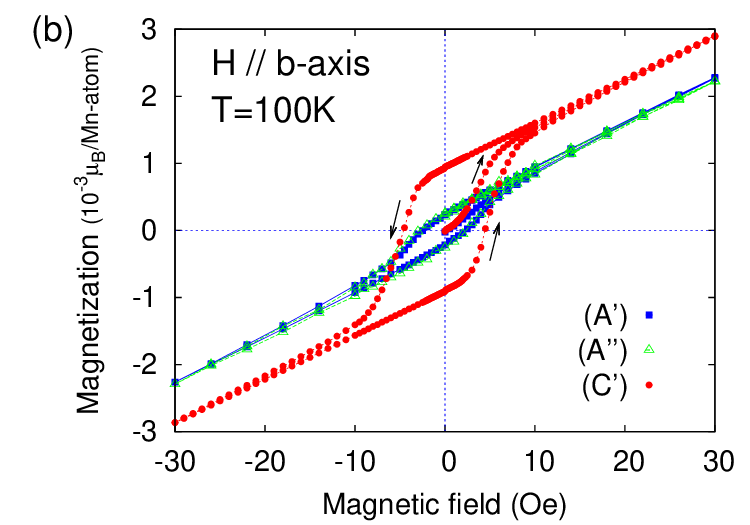}\\
\end{center}
\includegraphics[width=75mm]{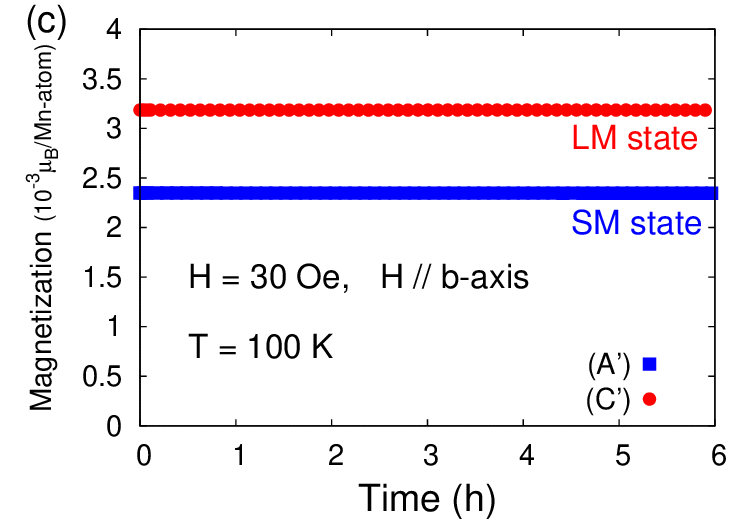}
\includegraphics[width=75mm]{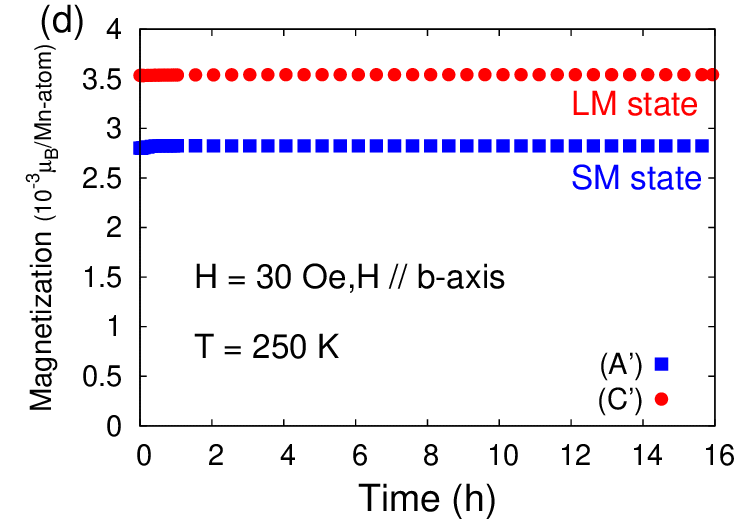}
\caption{(Color online) 
(a) Schematic diagram of the measurement processes (A'), (A''), and (C').
The solid circles represent the measurement temperatures of 100\,K (or 250\,K).
The dotted arrows represent approaching procedures for the measurement temperatures.
A magnetic field of 10\,kOe was applied at 350\,K  in the process (A'') (cross). 
(b) Field ($H$) dependences of the magnetization along the $b$-axis measured at $T = $ 100\,K via the processes (A'), (A''), and (C').
(c) and (d) Time evolution of the magnetizations along the $b$-axis at $H = 30$\,Oe; 
 the magnetic field was applied after the measurement temperature was reached. 
The measurement temperatures are (c) $100$\,K and (d) $250$\,K. 
}
\label{fig:M-Time-b1}
\end{figure}

Bearing in mind the striking temperature hysteresis shown in Fig.~\ref{fig:b-axis}(a), 
 we have performed $M-H$ measurements at fixed temperatures in the IT phase after three different processes.
The sample was set to 100~K via the following processes:
(A') the sample was cooled from 350 to 100\,K at zero field, 
(A'') the sample was cooled from 350 to 100\,K at a field of 10\,kOe, 
and (C') the sample was warmed from 5 to 100\,K after being cooled from 350 to 5\,K once at zero field. 
The sample has undergone the helimagnetic state once in the process (C'), corresponding to the $T$ process (C).
On the other hand, the processes (A') and (A'') correspond to the $T$ process (A). 
The processes are schematically shown in Fig.~\ref{fig:M-Time-b1}(a).
The measurements were performed by changing the field in the sequence of 0\,Oe $\rightarrow$ 30\,Oe $\rightarrow -30$\,Oe $\rightarrow$ 30\,Oe.
Figure \ref{fig:M-Time-b1}(b) shows the field ($H$) dependences of the magnetization along the $b$-axis at $T = 100$\,K in the three above-mentioned processes.
Ferromagnetic hysteresis loops were observed in all the processes.
This indicates that the IT phase  is a ferromagnetic phase with a very small spontaneous magnetization along the $b$-axis.
The spontaneous magnetization $M_{\mbox{\scriptsize s}}^{\parallel {\mbox{\scriptsize b}}}$ and the coercive field $H_{\mbox{\scriptsize c}}^{\parallel {\mbox{\scriptsize b}}}$ in the process (C') are larger than those in the processes (A') and (A'').
In particular, $M_{\mbox{\scriptsize s}}^{\parallel {\mbox{\scriptsize b}}}$ is enhanced five times.
The $M_{\mbox{\scriptsize s}}^{\parallel {\mbox{\scriptsize b}}}$ values in the processes (A') and (C') are about 2 $\times 10^{-4}\,\mb$/Mn atom and 1 $\times 10^{-3}\,\mb$/Mn atom, respectively.
This process-dependent behavior of $M_{\mbox{\scriptsize s}}^{\parallel {\mbox{\scriptsize b}}}$ corresponds to that observed in the $T$ dependence of the magnetization shown in Fig.~\ref{fig:b-axis}(a).

The results of the magnetization measurements indicate a strong correlation between the LT and IT phases
 and the presence of two different magnetic states in the IT phase for $\TN < T < T^{\ast}$:
 a large-magnetization (LM) state and a small-magnetization (SM) state. 
The LM state is realized only by cooling the sample once to the LT phase.
On the other hand, we consider that the SM state is restabilized after the sample is warmed up to the paramagnetic temperature ($T > \Tc$),
  because the reproducibility of the hysteresis behavior was confirmed as long as the sample was warmed up to 350\,K before the measurements.
One might consider that the large magnetic moment of the LM state can also be induced by strong magnetic field even without cooling the sample below $\TN$.
However, the magnetization curve for the process (A'') excellently traces that for the process (A').
This clearly indicates that a magnetic field of 10\,kOe, which is much higher than the coercive field of about 5\,Oe, cannot stabilized the LM state.
Only the $T$ process undergoing the LT phase can induce the LM state in the IT phase.

In order to determine which of the two states is in thermodynamic equilibrium,
we measured the time evolution of the magnetizations in the LM and SM states.
Figure \ref{fig:M-Time-b1}(c) shows the time dependences of the magnetizations at $H = 30$~Oe in the processes (A') and (C'); the magnetic field was applied after the measurement temperature was reached.
The magnetic states measured in the processes (A') and (C') correspond to the SM and LM states, respectively.
Both magnetizations exhibit no time dependence up to 6 h.
We also performed the same measurement at a higher temperature $T = 250$~K, where the thermal relaxation of the magnetization should be much faster than that at 100\,K.
Neverthless, no time dependences were found up to 16 h, as shown in Fig. \ref{fig:M-Time-b1}(d).
These results indicate that the energy barrier between the two states is so high that the relaxation does not occur in an experimentally accessible time scale.
Therefore, we cannot conclude which of the two states is the thermodynamically equilibrium state.

\begin{figure}[h]
\includegraphics[width=75mm]{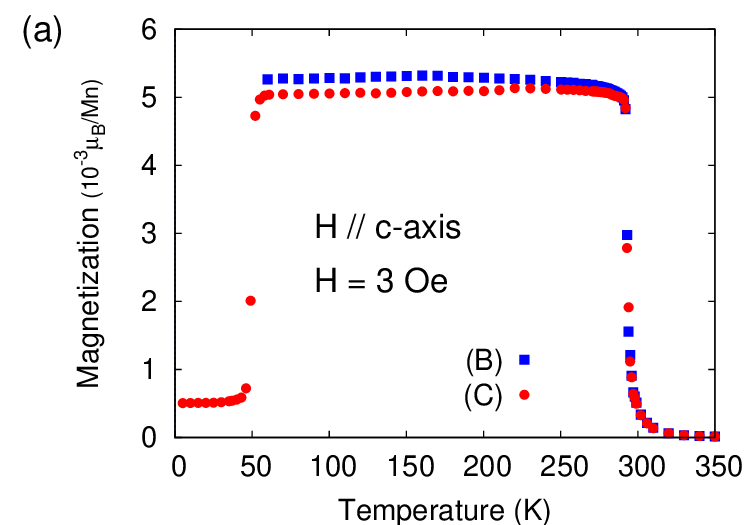}
\includegraphics[width=75mm]{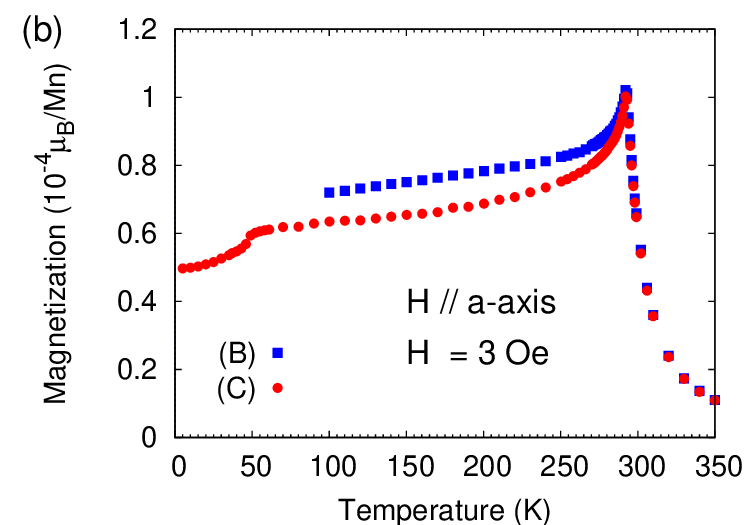}
\caption{\label{fig:ac-axis}(Color online) 
Temperature dependences of the magnetizations along the (a) $c$-axis and (b) $a$-axis.
In both figures, the data measured in the two processes (B) and (C) are shown. 
The processes (B) and (C) are warm-up processes after the sample was cooled down to 55 or 100\,K ($>\TN$) and 5\,K ($<\TN$), respectively.
}
\begin{center}
\includegraphics[width=75mm]{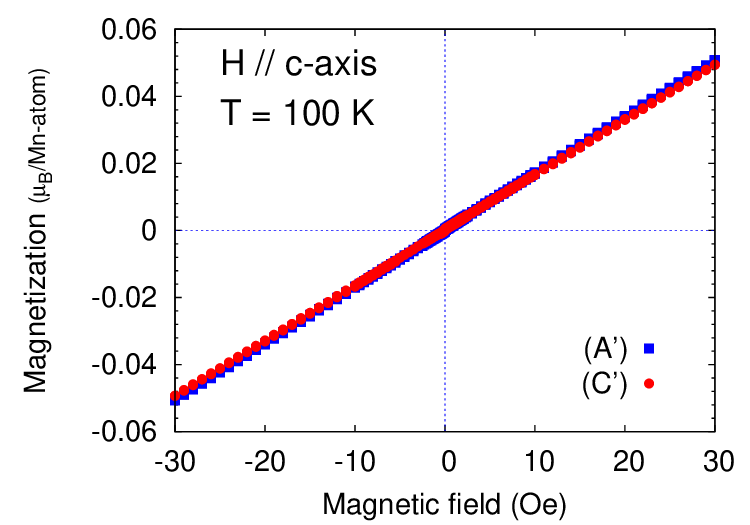}
\caption{\label{fig:M-H-c-axis}(Color online) 
Magnetic field dependences of the magnetization along the $c$-axis measured at $T =$100\,K  via the following processes: 
 (A') the sample was cooled from 350 to 100\,K at zero field and
 (C') the sample was warmed from 5 to 100\,K after being cooled from 350 to 5\,K once at zero field. 
}
\end{center}
\end{figure}

Figures \ref{fig:ac-axis}(a) and \ref{fig:ac-axis}(b) show the temperature dependences of the magnetizations along the $c$- and $a$-axes, respectively.
The data were obtained in the processes (B) and (C) mentioned earlier;
process-dependent behaviors were also observed along both the directions below $\Tc$. 
The magnetizations along the $c$- and $a$-axes are suppressed once the sample is cooled down to the helimagnetic phase, in contrast to that along the $b$-axis.
Figure~\ref{fig:M-H-c-axis} shows the magnetization curve along the $c$-axis measured at $T=$ 100~K after approaching the measurement temperature in the processes (A') and (C').
An ultrasoft ferromagnetic behavior was observed.
The coercive field and remnant magnetization along the $c$-axis are zero within the error margin.

\subsection{\label{sec:level5}Neutron scattering experiments}

\begin{figure}[t]
\includegraphics[width=74mm]{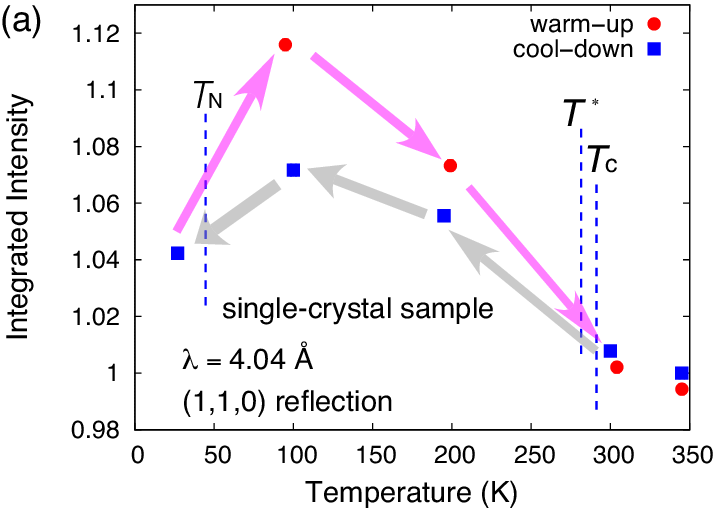}
\includegraphics[width=80mm]{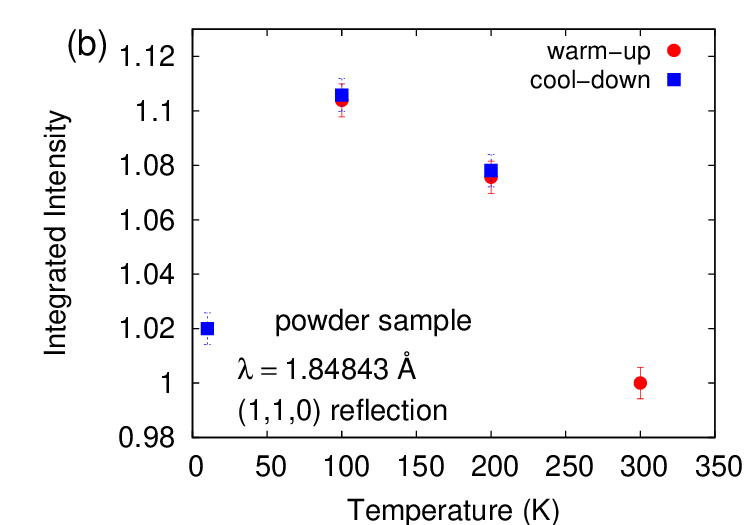}
\caption{\label{fig:110}(Color online) 
$T$ dependences of the normalized integrated intensities of the (1,1,0) reflection obtained using (a) single-crystal and (b) powder samples.
In both figures, the data obtained in the cool-down and warm-up processes are shown. 
The thick arrows represent the sequence of measurement temperatures.
The error bars are less than the point size in (a).
}
\end{figure}


In order to investigate the unusual temperature hysteresis observed in the magnetization measurements in more detail,
 we measured the temperature dependence of nuclear Bragg reflections by neutron scattering experiments using single-crystalline and powder samples.
Two $T$ processes were employed in the measurements: a warm-up process from 9\,K and a cool-down process from 350\,K, corresponding to the $T$ processes (A) and (C), respectively. 
Figure \ref{fig:110}(a) shows the temperature dependence of the integrated intensity of the (1,1,0) reflection from the single crystalline sample.
In the figure, each intensity is normalized by the value at $T = 350$\,K in the cool-down process.
The thick arrows in the figure represent the directions of the temperature shift of the measurements.
The intensity in the IT phase ($\TN < T < T^{\ast}$) measured in the warm-up process is considerably larger than that in the cool-down process.
One might consider that the hysteresis of the $(1,1,0)$ reflection intensity originates from the hysteresis of the ferromagnetic magnetization, which was indeed detected in the magnetization study,
 because the (1,1,0) reflection has a magnetic contribution due to the ferromagnetic ordering in the IT and HT phases.
However, the difference between the magnetizations for the cool-down and warm-up processes is a few $10^{-3}\,\mb$, which is too small to account for the 5\,\% change in the (1,1,0) reflection intensity detected in the neutron scattering experiments. 
The temperature dependence of the (1,1,0) reflection intensity from the powder sample is shown in Fig.~\ref{fig:110}(b).
No difference between the intensities in the two $T$ processes is found in the powder experiment.
The hysteresis behavior observed only in the single-crystalline sample will be discussed in Sect. \ref{sec:level7}.

\begin{figure}[t]
\hspace{2cm}
\includegraphics[width=120mm]{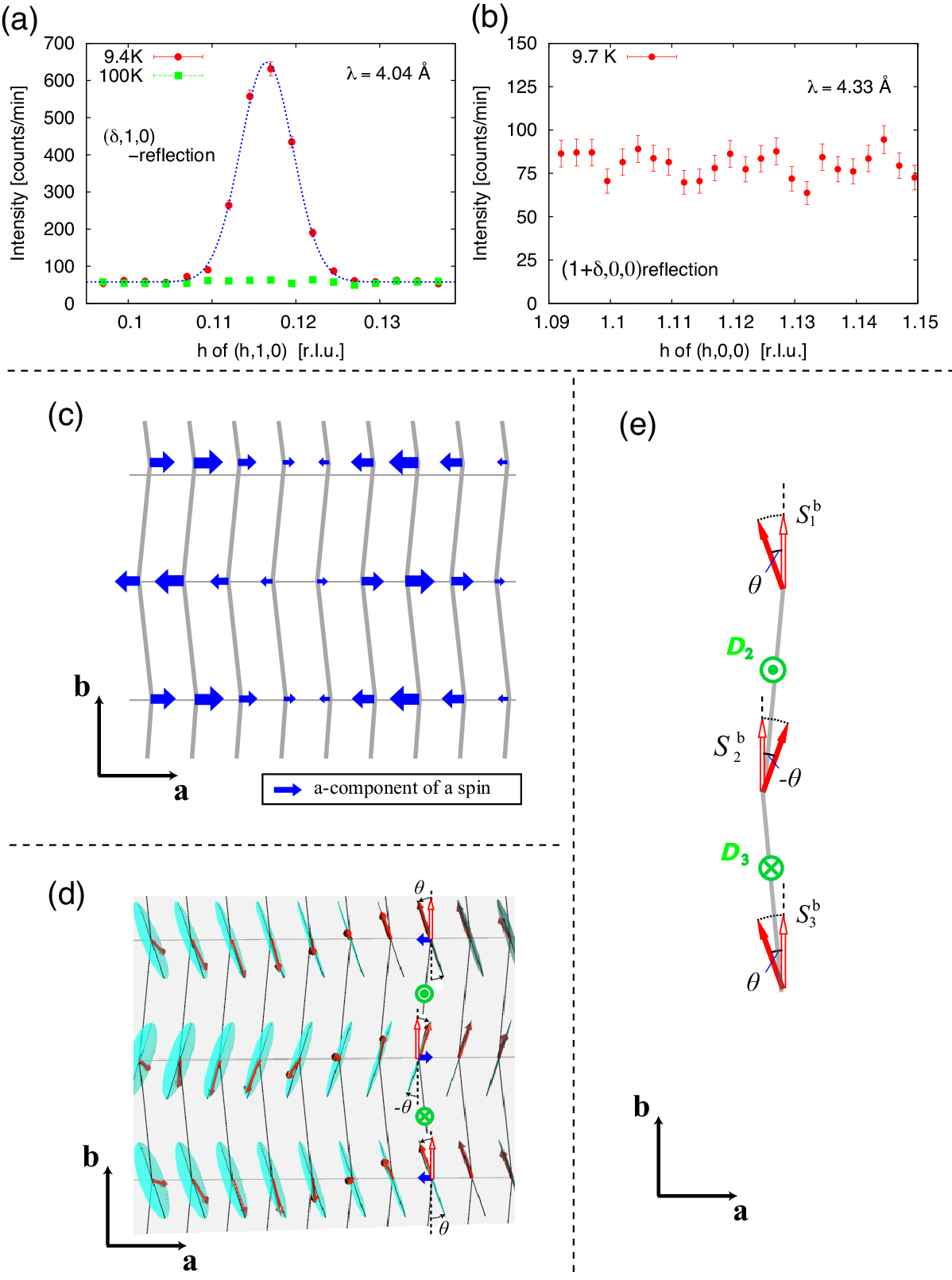}
\caption{\label{fig:d10}(Color online) 
(a) Neutron scattering profiles of the newly observed magnetic reflection at ($\delta$,1,0).
      The data at $T = 9.4$\,K ($< \TN$) and $T = 100$\,K ($> \TN$) are shown.
(b) Neutron scattering profile at (1+$\delta$,0,0) at $T = 9.7$\,K.
(c) Schematic illustration of the ($\delta$,1,0) modulation of the a-component of the Mn spins.
	For clarity, the drawn a-components are enlarged from the scale of those shown in (d). 
(d) Schematic illustration of the tilted helimagnetic structure, obtained by superposing the ($\delta$,1,0) modulation shown in (c) on the simple proper helimagnetic structure. 
      The cross and dot marks represent the DM vectors at the centers of nn Mn sites directed to the c- and anti-c-directions alternately, respectively.
(e) Schematic view of the possible explanation of the alternate tilt of the helical plane from the $bc$-plane to the a-direction by the DM interaction.
      One nn Mn chain is selected from this figure.
      The DM vectors at the center of the Mn pairs are represented in the same manner as in (d). 
}
\end{figure}

We also performed single-crystal neutron scattering experiments in the LT phase and found a new magnetic Bragg reflection at the ($\delta$,1,0) position, as shown in Fig. \ref{fig:d10}(a).
The magnetic propagation vector of the helimagnetic state of MnP reported previously is ($\delta$,0,0), and the (0,1,0) reflection is forbidden. Thus, the magnetic ($\delta$,1,0) reflection is fundamentally forbidden.
In reality, the previously reported magnetic structure, a double spiral structure in which Mn spins in the chemical unit cell have different rotating phases\cite{Forsyth1966,Obara1980}, allows for a very weak ($\delta$,1,0) reflection.
Nonetheless, it is 70 times smaller than the observed intensity of the ($\delta$,1,0) reflection.
This large discrepancy clearly requires reconsideration of the magnetic structure in the helimagnetic state: an additional modulation with the propagation vector of ($\delta$,1,0) should be considered.
Note that the reflection at (1+$\delta$,0,0) = ($\delta$,1,0) + (1,-1,0) is absent, as shown in Fig. \ref{fig:d10}(b).
This indicates that the modulation with ($\delta$,1,0) has only the a-axis component.
This additional modulation is shown schematically in Fig. \ref{fig:d10}(c).
By superimposing this modulation to the fundamental proper helimagnetic structure with ${\bm Q} =$ ($\delta$,0,0), the magnetic structure shown in Fig. \ref{fig:d10}(d) is obtained.
In this structure, the helical planes are tilted to the a-direction from the $bc$-plane by angles of $\theta$ and $-\theta$, alternately along the $b$-axis.
The size of the a-component $m^{\mbox{\scriptsize a}}$ and the tilt angle $\theta$ are estimated to be about 0.046\,$\mb$/Mn atom and 2.0 deg from the integrated intensity of the
($\delta$,1,0) reflection, respectively.
\clearpage

\section{\label{sec:level6}Discussion}

\subsection{\label{sec:level7} Origin of the tilted helimagnetic structure}
First, we discuss the origin of the newly proposed magnetic structure below $\TN$, i.e., the alternately tilted helimagnetic structure described in Sect. 3.2,
 from the viewpoint of the crystal symmetry of MnP. 
As described in Sect. 1, inversion symmetry is absent at the center of nn Mn bonds, and thus a finite DM interaction is allowed for them.
The DM vector ${\bm D}$ should be perpendicular to the $ab$-plane because the $ab$-plane, including nn Mn sites, is a mirror plane [see Fig.~\ref{fig:crystal-1}(a)].
As shown in Fig.~\ref{fig:crystal-1}(b), Mn atoms form a zigzag chain along the b-direction by connecting nn Mn sites (nn Mn chain).
The DM vectors directed parallel and antiparallel to the $c$-axis are aligned alternately along this chain
  because the crystal structure is invariant under a glide operation along ${\bm b}$.

Considering of such alternate DM interactions between nn Mn pairs, the alternate tilt of the helical plane can be naturally elucidated.
The DM interaction with ${\bm D} \parallel c$ only acts on Mn spins projected to the ab-plane.
In the fundamental helimagnetic structure proposed earlier, the projected Mn spins are parallel to the $b$-axis and their directions in a nn Mn chain are the same [see open arrows in Fig.~\ref{fig:d10}(e)].
The DM interaction with the alternate nn DM vector ${\bm D}_{i} = (0,0,(-1)^{i}D_{\mbox{\scriptsize s}}^{\mbox{\scriptsize c}})$ tilts Mn spins in the nn Mn chain toward the a-direction with the alternate angles $\theta$ and $-\theta$ [see filled arrows in Fig.~\ref{fig:d10}(e)].
The magnitude of the tilt angle $\theta$ is determined by the ratio of the strength of the $c$-component of the DM interaction $D_{\mbox{\scriptsize s}}^{\mbox{\scriptsize c}}$ and the symmetric exchange interactions $J$.
Consequently, the alternately tilted helimagnetic structure shown in Fig.~\ref{fig:d10}(d) is realized. 
From the experimentally obtained tilt angle of $\theta = 2.0$ deg, the ratio of the strength of the DM interaction to the symmetric exchange interaction $D_{\mbox{\scriptsize s}}^{\mbox{\scriptsize c}}/J$ is estimated to be about 0.03.
Very recently, Shiomi \textit{et al.} have performed Hall resistivity measurements of MnP~\cite{shiomi2012emergence} and
  observed topological Hall resistivity in a fan like phase at high magnetic fields and low temperatures, suggestive of a non-coplanar spin structure.
Their results are in agreement with the tilted helimagnetic structure at zero field proposed in this article.

\subsection{\label{sec:level7} Magnetic structure and temperature hysteresis in the IT phase}

Next, we discuss the magnetic structure and temperature hysteresis in the IT phase.
In our measurements, a weak ferromagnetic behavior along the $b$-axis was observed with a finite coercivity at $\TN < T < T^*$.
On the other hand, an ultrasoft ferromagnetic behavior was observed in the magnetization curve along the $c$-axis, \textit{i.e.,} the spontaneous magnetization and coercivity were not found [see Fig.~\ref{fig:M-H-c-axis}].
The most striking feature of MnP found in this study is the unusual temperature hysteresis:
 (i) the enhancement of the spontaneous magnetization along the $b$-axis [Figs.~\ref{fig:b-axis}(a), \ref{fig:M-Time-b1}, and \ref{fig:LM-SM-states}], and
 (ii) the enhancement of the Bragg intensity in the neutron scattering measurement using the single crystal [Fig.~\ref{fig:110}(a)] are observed in the IT phase only after the sample is cooled down to the LT phase.
If a simple canted ferromagnetic structure expected before is realized in the IT phase\,\cite{becerra2000a}, it is difficult to explain the above results.
Then, what is the real magnetic structure in the IT phase?
What is the origin of the unusual temperature hysteresis?

\begin{figure}[h]
\begin{center}
\includegraphics[width=100mm]{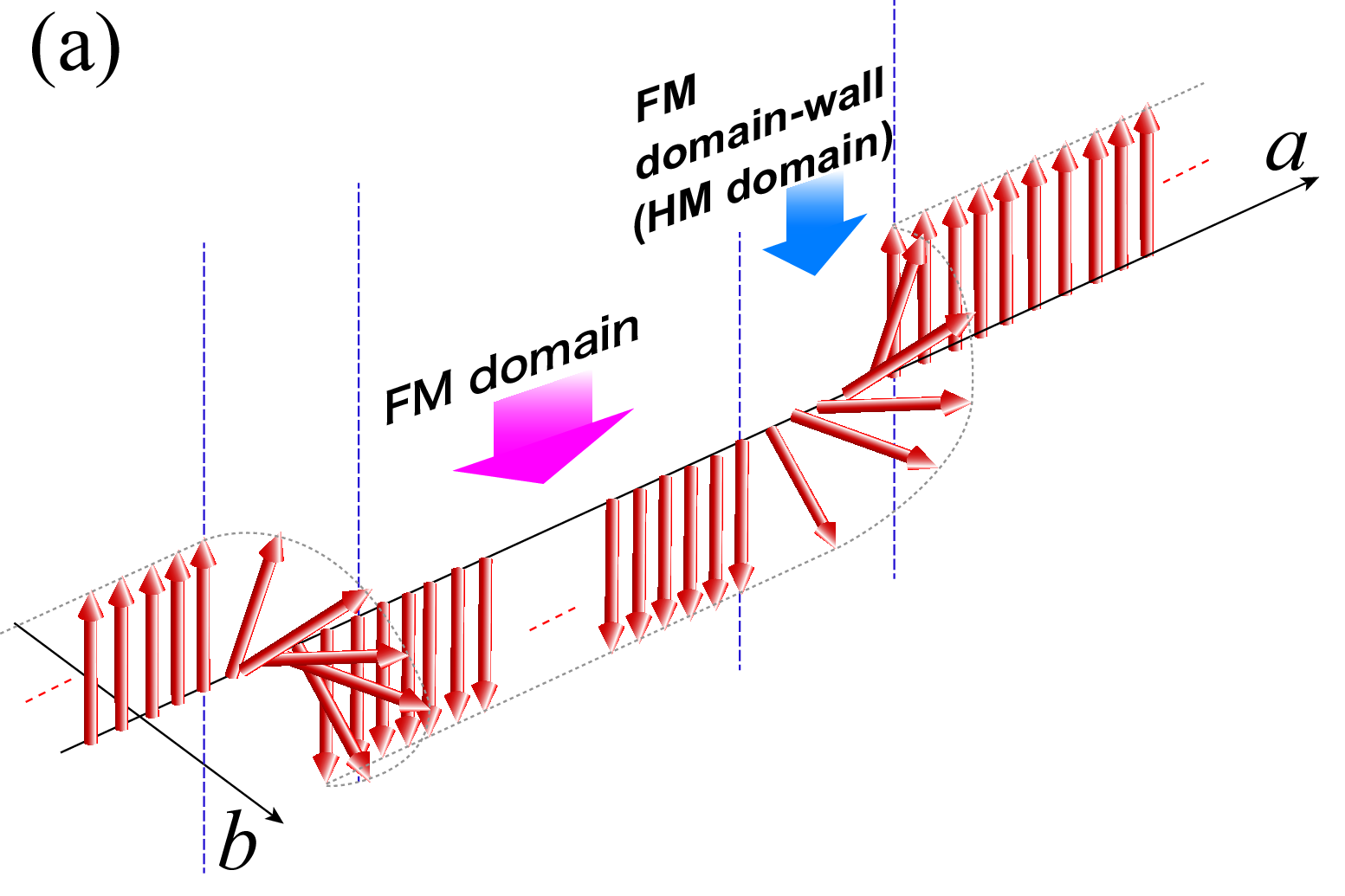}
\includegraphics[width=100mm]{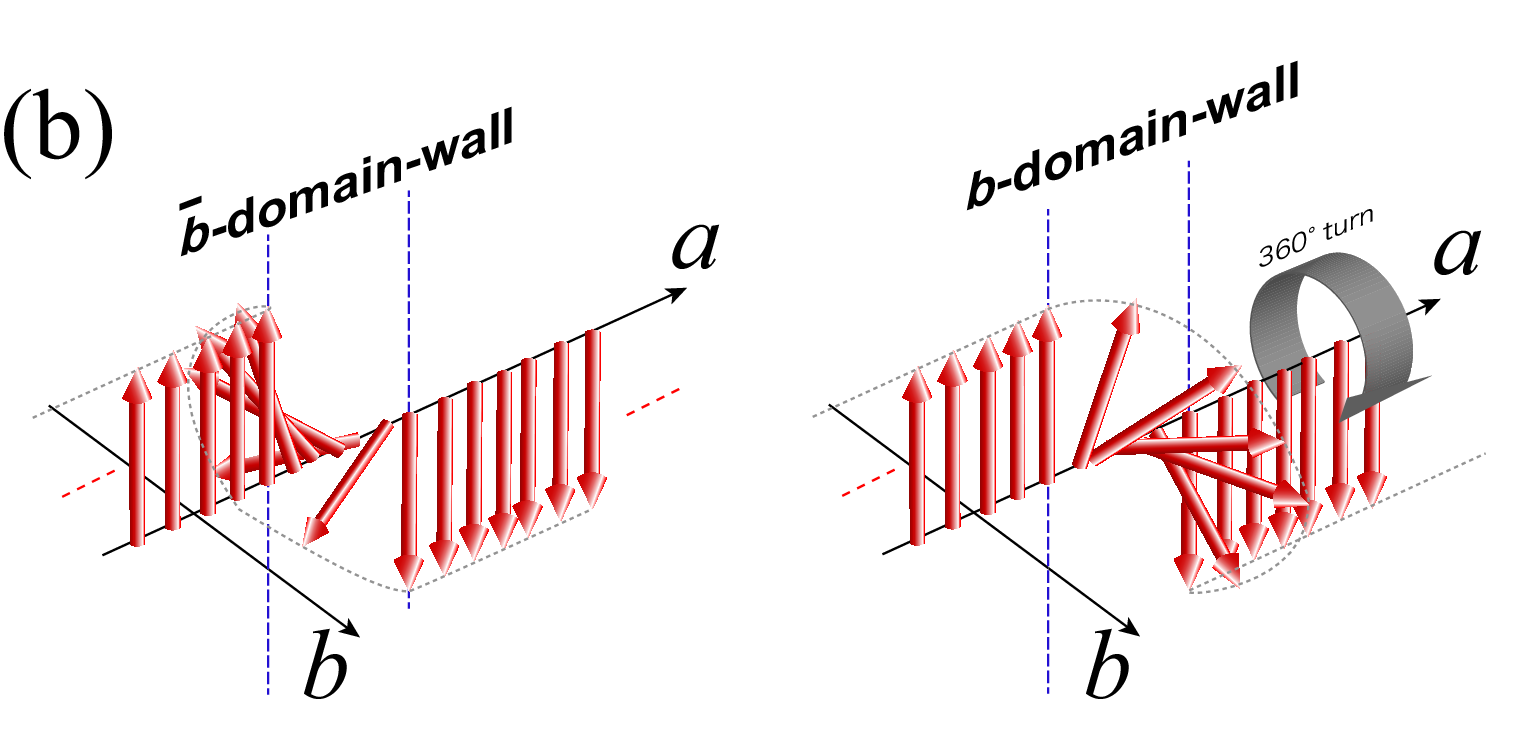}
\includegraphics[width=100mm]{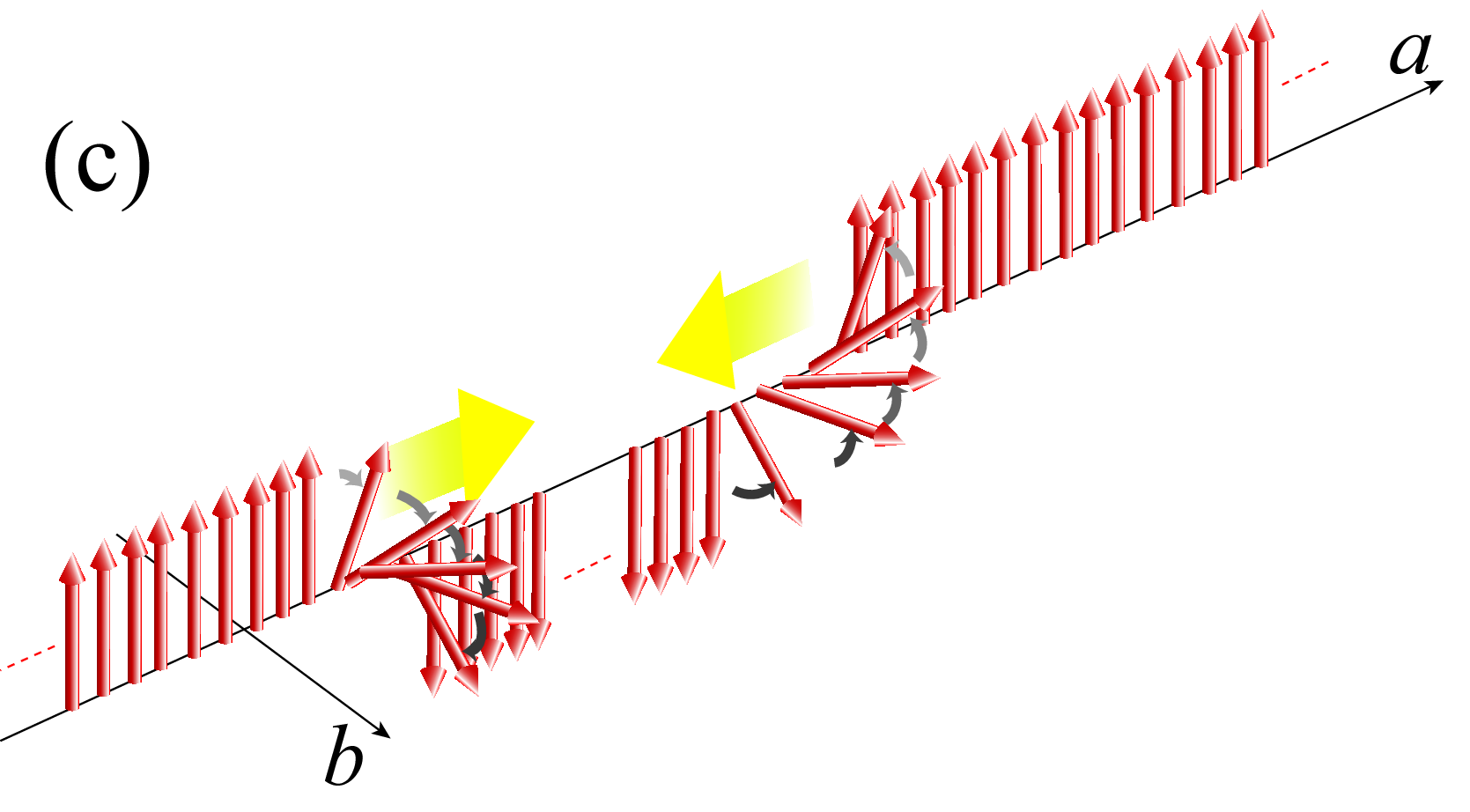}
\caption{(Color online) (a) Schematic illustration of the long periodic striped ferromagnetic domain structure observed using  Lorentz transmission electron microscopy and small-angle electron diffraction analysis in the IT phase ~\cite{Koyama2012}. The turn directions of Mn spins, i.e., the vector chirality, in ferromagnetic (FM) domain-walls are alternately arranged along the $a$-axis. 
A Bloch-type FM domain-wall can be regarded as a LT helimagnetic (HM) domain.
(b) Schematic illustrations of $\bar{b}$- and $b$-domain-walls.
 When a magnetic field along the $b$-direction is applied,  discontinuous turns of the spins in the domain-wall towards the mirror-symmetrical position,  or 360 deg turns of the ferromagnetic $c$-domain are necessary to change from a $\bar{b}$-domain-wall to a $b$-domain-wall.
(c) Schematic illustration of the continuous change in the spin configuration induced by applying magnetic field along the c-direction.
}
\label{fig:kink-crystal} 
\end{center}
\end{figure}

\begin{figure}[h]
\begin{center}
\includegraphics[width=120mm]{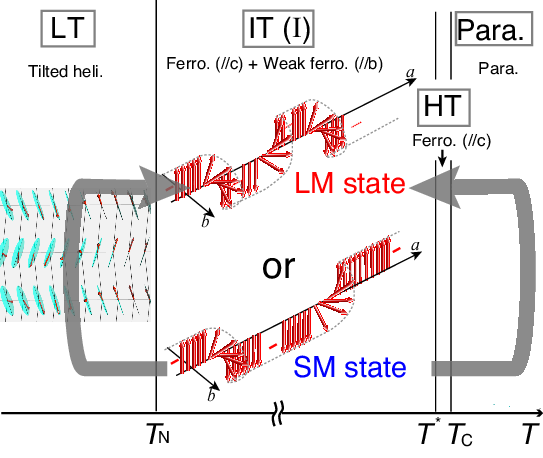}
\caption{(Color online) 
Schematic diagram of the temperature hysteresis in MnP.
In the IT phase, the LM state  is realized only after the sample is cooled down to the LT helimagnetic phase.
Additional LT helimagnetic domains (FM domain-walls) may be inserted in the IT phase after the sample is cooled down to the LT phase.}
\label{fig:LM-SM-states} 
\end{center}
\end{figure}

Recently, Koyama et al. have observed a striped ferromagnetic domain structure at 120 K,
 corresponding to the temperature region of the IT phase, using Lorentz transmission microscopy and small-angle electron diffraction analysis for thin single-crystal MnP \cite{Koyama2012}.
In this structure, ferromagnetic domains 
 with a length of $\sim$ 64 nm are periodically aligned, with Bloch-type domain-walls along the $a$-axis, as shown in Fig.~\ref{fig:kink-crystal}(a). 
The weak ferromagnetic behavior along the $b$-axis in the IT phase, observed in our experiments, 
 can be easily understood on the basis of this striped ferromagnetic domain structure. 
In a Bloch-type domain-wall, Mn spins turn by 180 deg from the $c$-direction to the $\bar{c}$-direction, 
or oppositely. The domain-wall has a finite magnetic moment toward the $b$- or $\bar{b}$-direction, 
which can be responsible for the weak ferromagnetic behavior in  the IT phase, 
and  the ferromagnetic $M-H$ curves shown in Fig. \ref{fig:M-Time-b1} (b) can be explained as follows. 
By cooling at zero field, corresponding to the initial state of the process (A') or (C'), 
the domain-walls with magnetic moments along the $b$- and $\bar{b}$-directions 
($b$- and $\bar{b}$-domain-walls, respectively) equally coexist. 
When magnetic field along the $b$-direction is applied, 
the magnetic moments of $\bar{b}$-domain-walls are flipped to the $b$-direction. 
Such a flip is accompanied by discontinuous turns of the spins in the domain-walls towards the mirror-symmetrical position, or by 360 deg turns of ferromagnetic $c$-domains, therefore a finite coercive field is needed to overcome the gap energy due to the magnetic anisotropy [see Fig. \ref{fig:kink-crystal}(b)].
In the "saturated"-region, out of the hysteresis loop in Fig. 3(b), all domain-walls are $b$-domain-walls  
and the turn directions of Mn spins in domain-walls alternately aline. Indeed, Koyama \textit{et al.} found 
an alternate arrangement of the turn direction of Mn spins, the vector chirality, in Bloch-type domain-walls.
The striped domain structure is near a ferromagnetic structure, but it is essentially an antiferromagnetic long-range order.
Therefore, the ultrasoft ferromagnetic behavior along the $c$-axis shown in Fig. \ref{fig:M-H-c-axis} can also be simply interpreted as an antiferromagnetic behavior for this magnetic structure as follows.
When the magnetic field along the $c$-direction is applied, the magnetic moments of ferromagnetic $\bar{c}$-domains turn to the ${c}$-direction easily and continuously because this process can be induced by movements of domain-walls with continuous turning of the spins in  domain-walls, as shown in Fig. \ref{fig:kink-crystal}(c).
When the magnetic field is returned to zero, the magnetic structure continuously returns to the striped ferromagnetic domain one without coercive field.

The unusual temperature hysteresis in the IT phase can also be understood on the basis of the striped domain structure. 
Here, we assume that the turn pitch of Mn spins in Bloch-type domain-walls is equal to that of the LT helimagnetic state. 
In other words, we speculate that Bloch-type domain-walls of the striped structure in the IT phase is equal to the LT helimagnetic domain with a length equal to its half-period [see Fig.\,\ref{fig:kink-crystal}(a)]. 
On the basis of this assumption, 
 some parts of the LT helimagnetic structure may remain as additional LT helimagnetic domains even in the IT phase after the sample has undergone to the LT phase.
The enhancement of the magnetization along the b-axis in the IT phase after undergoing the LT phase can be 
 explained by the additional insertion of LT helimagnetic domains, as shown in Fig.\,\ref{fig:LM-SM-states}, which are the seeds of ferromagnetism along the $b$-axis. 
Another hysteresis behavior, the increase in the neutron Bragg scattering intensity only in the single crystal, 
 can be explained as a reduction of the secondary extinction effect. 
If the mosaicity of crystal is very small, 
 the secondary extinction effect reduces the Bragg scattering intensity. 
Additional insertion of helimagnetic domains by entering to the IT phase from the LT phase is expected to induce a higher strain in the crystal, 
 which enlarges the mosaicity of the crystal and suppresses the secondary extinction effect. 
Consequently, the Bragg scattering intensity is increased. 

What is a driving force for this long-periodic striped structure? Recently a similar magnetic structure,  
a chiral magnetic soliton lattice, has been found in the chiral magnet Cr$_{1/3}$NbS$_{2}$ \cite{togawa2012chiral}.  
Cr$_{1/3}$NbS$_{2}$ exhibits a helimagnetic structure stabilized by competition of the ferromagnetic coupling 
and DM interaction with the DM vector parallel to the helical propagation vector at zero field, 
 and a chiral magnetic soliton lattice is realized by applying magnetic field. 
This is not the case for MnP. 
In MnP, as discussed before, the DM vector is along the c-axis, being perpendicular to the propagation vector 
of the striped structure, and the striped structure is realized even at zero field. 

Note that the hysteresis behavior was observed in the entire temperature range of the IT phase, 
indicating that the IT phase is a nonequilibrium phase.
Hence, the long periodic striped structure 
can be driven by nonequilibrium effects. 
The proposed magnetic structure in the IT phase is recognized as 
 a striped structure consisting of alternately arranged HT ferromagnetic and LT helimagneic domains. 
If the HT ferromagnetic and LT helimagnetic states nearly degenerate in energy, the phase separation 
and self organization to form a striped structure can occur between two phases. 
Similar phenomena have been observed in strongly correlated electron systems where multistates compete with each other, 
such as in high-Tc cuprates \cite{pan2001microscopic,lang2002imaging} and manganites \cite{dagotto2005complexity,koyama2012ferromagnetic} 
and in many nonequilibrium systems \cite{cross1993pattern}.  
Further investigation to clarify the mechanism of the spontaneous formation of a striped structure in MnP 
 is interesting from the viewpoint of not only magnetic materials science but also nonequilibrium physics. 

In the Lorentz transmission electron microscopy and small-angle electron diffraction analysis in Ref.\,18,  thin-film samples of MnP were used for observation of the periodic magnetic domains.
In those thin-film samples, the periodic domain structure may be produced by the effects of the two-dimensionality or those of the surface.
If this is the case, the magnetic structure in the IT phase for bulk samples may be different from the long periodic striped structure.
Consequently, we need to discuss another scenario for the temperature hysteresis.
One possible structure is a canted antiferromagnetic structure.
Our magnetization measurement along the $a$-axis shows an antiferromagnetic behavior [Fig.~\ref{fig:ac-axis}(a)].
If the antiferromagnetic component along the a-axis exists in the IT phase, the DM interaction discussed in Sect. 4.1 
 tilts it to the b-axis, forming a canted antiferromagnetic structure with a weak ferromagnetic b-component. 
In this model, the temperature hysteresis can be interpreted by the inverse effect of the DM interaction.
If a lattice distortion that enhances the DM interaction occurs in the LT phase and remains even after the sample is warmed up to the IT phase,
 the $b$-component of the canted antiferromagnetic structure can be enhanced.
Such a lattice distortion, however, has not been observed directly until now.

\section{\label{sec:level9}Conclusions}

In summary, we have investigated the magnetic properties of MnP by magnetization and neutron scattering experiments.
The previously reported magnetic structures of MnP should be updated to the alternately tilted helimagnetic structure stabilized by the DM interaction for the LT phase.
Moreover, a weak ferromagnetic behavior along the $b$-axis was observed in the IT phase.
Surprisingly, quite a unusual temperature hysteresis for $b$-axis magnetization was also observed.
These behaviors in the IT phase can be understood by assuming a spontaneous formation of a stripe structure consisting of alternately arranged HT ferromagnetic and LT helimagnetic domains.

\section*{\label{sec:level10} Acknowledgements}

This work was partly supported by a Grant-in-Aid for the Global COE Program, "International Center for Integrated Research and Advanced Education in Material Science" and by a Grant-in-Aid for Scientific Research on Priority Areas "Novel States of Matter Induced by Frustration" (19052003) from the Ministry of Education, Culture, Sports, Science and Technology of Japan.


\begin{thebibliography}{10}

\bibitem{Huber1964}
E.~E. Huber and D.~H. Ridgley, Phys. Rev. {\bfseries 135},  1033 (1964).

\bibitem{Forsyth1966}
J.~Forsyth, S.~Pickart, and P.~Brown, Proc. Phys. Soc. {\bfseries 88},  333
  (1966).

\bibitem{Felcher1966}
G.~Felcher, J. Appl. Phys. {\bfseries 37},  1056 (1966).

\bibitem{komatsubara1974}
T.~Komatsubara, A.~Ishizaki, S.~Kusaka, and E.~Hirahara, Solid State Commun.
  {\bfseries 14},  741 (1974).

\bibitem{Obara1980}
H.~Obara, Y.~Endoh, Y.~Ishikawa, and T.~Komatsubara, J. Phys. Soc. Jpn.
  {\bfseries 49},  928 (1980).

\bibitem{Moon1982}
M.~Moon, R, J. Appl. Phys. {\bfseries 53},  1956 (1982).

\bibitem{becerra1980}
C.~C. Becerra, Y.~Shapira, N.~F. Oliveira~Jr, and T.~Chang, Phys. Rev. Lett.
  {\bfseries 44},  1692 (1980).

\bibitem{shapira1981}
Y.~Shapira, C.~Becerra, N.~Oliveira~Jr, and T.~Chang, Phys. Rev. B {\bfseries
  24},  2780 (1981).

\bibitem{Yoshizawa1985}
H.~Yoshizawa, S.~Shapiro, and T.~Komatsubara, J. Phys. Soc. Jpn. {\bfseries
  54},  3084 (1985).

\bibitem{Bindilatti1989}
V.~Bindilatti, C.~C. Becerra, and N.~F. Oliveira~Jr, Phys. Rev. B {\bfseries
  40},  9412 (1989).

\bibitem{becerra2000}
C.~C. Becerra, V.~Bindilatti, and N.~F. Oliveira~Jr, Phys. Rev. B {\bfseries
  62},  8965 (2000).

\bibitem{Zieba2000}
A.~Zieba, M.~Slota, and M.~Kucharczyk, Phys. Rev. B {\bfseries 61},  3435
  (2000).

\bibitem{Reis2008}
M.~Reis, R.~Rubinger, N.~Sobolev, M.~Valente, K.~Yamada, K.~Sato, Y.~Todate,
  A.~Bouravleuv, P.~von Ranke, and S.~Gama, Phys. Rev. B {\bfseries 77},
  104439 (2008).

\bibitem{becerra2000a}
C.~C. Becerra, J. Phys.: Condens. Matter {\bfseries 12},  5889 (2000).

\bibitem{Yamazaki2010}
T.~Yamazaki, Y.~Tabata, T.~Waki, H.~Nakamura, M.~Matsuura, and N.~Aso, J.
  Phys.: Conf. Ser. {\bfseries 200},  032079 (2010).

\bibitem{Ohoyama1998}
K.~Ohoyama, T.~Kanouchi, K.~Nemoto, M.~Ohashi, T.~Kajitani, and Y.~Yamaguchi,
  Jpn. J. Appl. Phys. {\bfseries 37},  3319 (1998).

\bibitem{shiomi2012emergence}
Y.~Shiomi, S.~Iguchi, and Y.~Tokura, Phys. Rev. B {\bfseries 86},  180404
  (2012).

\bibitem{Koyama2012}
T.~Koyama, S.~Yano, Y.~Togawa, Y.~Kousaka, S.~Mori, K.~Inoue, J.~Kishine, and
  J.~Akimitsu, J. Phys. Soc. Jpn. {\bfseries 81},  043701 (2012).

\bibitem{togawa2012chiral}
Y.~Togawa, T.~Koyama, K.~Takayanagi, S.~Mori, Y.~Kousaka, J.~Akimitsu,
  S.~Nishihara, K.~Inoue, A.~Ovchinnikov, and J.~Kishine, Phy. Rev. Lett.
  {\bfseries 108},  107202 (2012).

\bibitem{pan2001microscopic}
S.~H. Pan, J.~O'Neal, R.~Badzey, C.~Chamon, H.~Ding, J.~Engelbrecht, Z.~Wang,
  H.~Eisaki, S.~Uchida, A.~Gupta, et~al., Nature {\bfseries 413},  282 (2001).

\bibitem{lang2002imaging}
K.~Lang, V.~Madhavan, J.~Hoffman, E.~Hudson, H.~Eisaki, S.~Uchida, and
  J.~Davis, Nature {\bfseries 415},  412 (2002).

\bibitem{dagotto2005complexity}
E.~Dagotto, Science {\bfseries 309},  257 (2005).

\bibitem{koyama2012ferromagnetic}
T.~Koyama, Y.~Togawa, K.~Takenaka, and S.~Mori, J. Appl. Phys. {\bfseries 111},
   07B104 (2012).

\bibitem{cross1993pattern}
M.~C. Cross and P.~C. Hohenberg, Rev. Mod. Phys. {\bfseries 65},  851 (1993).

\end{thebibliography}
\end{document}